\documentclass[aps,prl,twocolumn,groupedaddress]{revtex4}

\usepackage{color}
\usepackage{graphicx}
\usepackage{amssymb, amsmath, amsfonts, wasysym}

\begin{document}

\title{Optical gecko toe: Optically controlled attractive near-field forces between plasmonic metamaterials and dielectric or metal surfaces}

\author{J.~Zhang, K.~F.~MacDonald, and N.~I.~Zheludev}
\email{niz@orc.soton.ac.uk}
\homepage{www.nanophotonics.org.uk/niz}
\affiliation{Optoelectronics Research Centre \& Centre for Photonic Metamaterials, University of Southampton, Southampton, SO17 1BJ, UK}

\begin{abstract}
On the mesoscopic scale, electromagnetic forces are of fundamental importance to an enormously diverse range of systems, from optical tweezers to the adhesion of gecko toes. Here we show that a strong light-driven force may be generated when a plasmonic metamaterial is illuminated in close proximity to a dielectric or metal surface. This near-field force can exceed radiation pressure and Casimir forces to provide an optically controlled adhesion mechanism mimicking the gecko toe: at illumination intensities of just a few tens of nW/$\mathrm{\mu}$m$^2$ it is sufficient to overcome the Earth's gravitational pull.
\end{abstract}

\maketitle

In various guises, electromagnetic forces are extremely important in mesoscopic systems: They are exploited in all forms of optical tweezing, manipulation and binding~\cite{Dholakia2011,Padgett2011,Dholakia2010,Chen2011a}, in optomechanical photonic devices~\cite{Kippenberg2008,Li2008a,Ma2011,Roels2009,Wiederhecker2009,Taniyama2008,Liu2009}, and may offer a mechanism of `quantum lubrication' between surfaces\cite{Munday2009}; The fact that light exerts pressure upon any surface exposed to it (deduced theoretically by James Clerk Maxwell in 1871 and proven experimentally in 1901 by Lebedev~\cite{Lebedev1901} and in 1903 by Nichols \& Hull~\cite{Nichols1903}) can be exploited to generate `optical lift'\cite{SwartzlanderJr.2011}; And in nature, van der Waals interactions underpin the gecko's remarkable ability to overcome gravity by sticking to walls and ceilings~\cite{Autumn2002}. In recent years it has become clear that plasmonic systems can provide unprecedented control over optical fields on the nanoscale, offering gigantic field enhancement, subwavelength light localization and strongly enhanced interactions between nano-objects~\cite{Zhao2010, Liu2010b,Wang2011,Liu2011}. They present opportunities, for example, to extend/enhance the functionality of optical tweezers/traps~\cite{Juan2011} and to experimentally investigate the idea that structures with negative refractive indices will experience negative radiation pressure~\cite{Veselago1968,Lezec2011}.

Here we show that in addition to the conventional, well-understood force of radiation pressure a much stronger light-driven near-field force may be generated between an illuminated planar plasmonic metamaterial and a dielectric or metallic surface. This newly identified near-field force has a resonant nature linked to the excitation of the metamaterial's plasmonic mode and acts to close the gap between the metamaterial film and the surface (Fig.~1). This `optical adhesion' force exists alongside interfacial Casimir forces and can overcome both radiation pressure and, like the gecko toe, gravity.

\begin{figure}
\includegraphics[width=82mm]{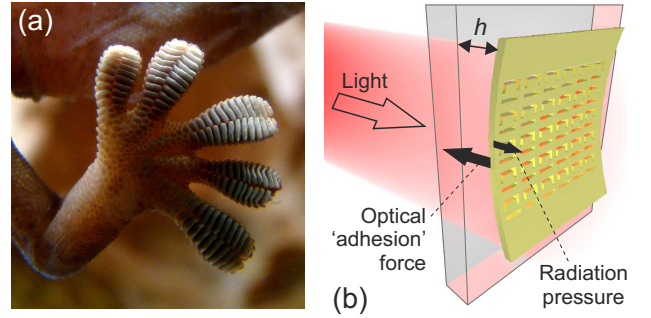}
\caption{\textbf{Gecko toes and their optical analogue. a,} Gecko toes sticking to a smooth glass wall~\cite{Torrissen}. \textbf{b,} Artistic impression of a metamaterial film attracted by a beam of light to a dielectric surface.}
\vspace{-10pt}
\end{figure}

Within the framework of classical electrodynamics the components of the total time-averaged force $\textbf{F}$ acting on a metamaterial structure illuminated with light can be calculated using a surface integral:

\begin{equation}
\langle F_i^{} \rangle = \displaystyle\oiint\limits_{S} \langle T_{ij}^{} \rangle \, n_j^{} \, \mathrm{d}S
\end{equation}

\noindent where $S$ is a bounding surface around the metamaterial and $T_{ij}$ is the time-averaged Maxwell stress tensor:

\begin{equation}
\begin{aligned}
\langle T_{ij} \rangle = \frac{1}{2} \mathrm{Re}\bigg[\varepsilon_0^{} &\bigg(E_i^{} E_j^\ast - \frac{1}{2} \displaystyle\sum\limits_{k} E_k^{} E_k^\ast\bigg)\\
+ \, \mu_0^{} &\bigg(H_i^{} H_j^\ast - \frac{1}{2} \sum_k H_k^{} H_k^\ast\bigg)\bigg]
\end{aligned}
\end{equation}

The stress tensor integral equation (1) encompasses both the radiation pressure and near-field force - the focus of this study. It does not include Casimir forces, which are derived from vacuum quantum fluctuation and thus exist even in the absence of illumination. (It will be demonstrated below that the classical electromagnetic force can be much stronger than the Casimir force.)

Radiation pressure arises through transfer of momentum between photons and any surface on which they impinge. It depends on the reflection $R$ and absorption $A$ coefficients of the surface according to the equation $F_r = (2R + A)P/c$, where $c$ is the speed of light in vacuum and $P$ is the power of the incident light, and assumes a maximum value of $2P/c$ when the reflectivity of a surface is 100$\%$.

In addition to this radiation force, a near-field optical force emerges when a metamaterial is placed in close proximity to another object and the evanescent field of the nanostructure encounters that object. The magnitude of this force depends on how strongly optical energy is trapped in the metamaterial and on the electromagnetic properties of the nearby object. In general, the higher the refractive index of the nearby object is, the stronger the interaction will be and in all cases studied here the near-field optical force is attractive. In what follows, the ratio $|\textbf{F}|:P/c$ is employed as a dimensionless measure of optical force~\cite{Rodriguez2011}.

In the present study, forces acting on metamaterial structures are evaluated via the Maxwell stress tensor integral Eq.~(1) with electric $E$ and magnetic $H$ field distributions obtained from fully three-dimensional finite element Maxwell solver simulations (Comsol MultiPhysics). This analysis utilizes established experimental values of the complex dielectric parameters for gold~\cite{Palik1984}, excludes losses in dielectric media and assumes normally incident, narrow-band coherent illumination. By modeling a single `meta-molecule' (translation unit cell of the metamaterial design) with periodic boundary conditions, calculations assumed a planar metamaterial array of infinite extent, parallel to the surface of a nearby semi-infinite metal or dielectric. The surface of integration $S$ was defined as a rectangular parallelepiped enclosing the meta-molecule, with walls along each of the four periodic boundaries and \emph{outside} each of the two free surfaces of the metamaterial film. The same numerical model provides data on the transmission $T$ and reflectance $R$ of the structure. In all cases, stress tensor results are verified against independent calculations of force based on the variation with gap size of the total energy trapped in the system~\cite{Povinelli2005}.

We first consider optical forces between a gold metamaterial film and the surface of a semi-infinite transparent dielectric (Fig.~2). The metamaterial here is taken to comprise a two-dimensional square array of asymmetric split rings - a popular design recognized for its strong light confinement and simplicity of fabrication~\cite{Tanaka2010,Nikolaenko2010}. Dimensional details are shown inset to Fig.~2a.

\begin{figure}
\includegraphics[width=82mm]{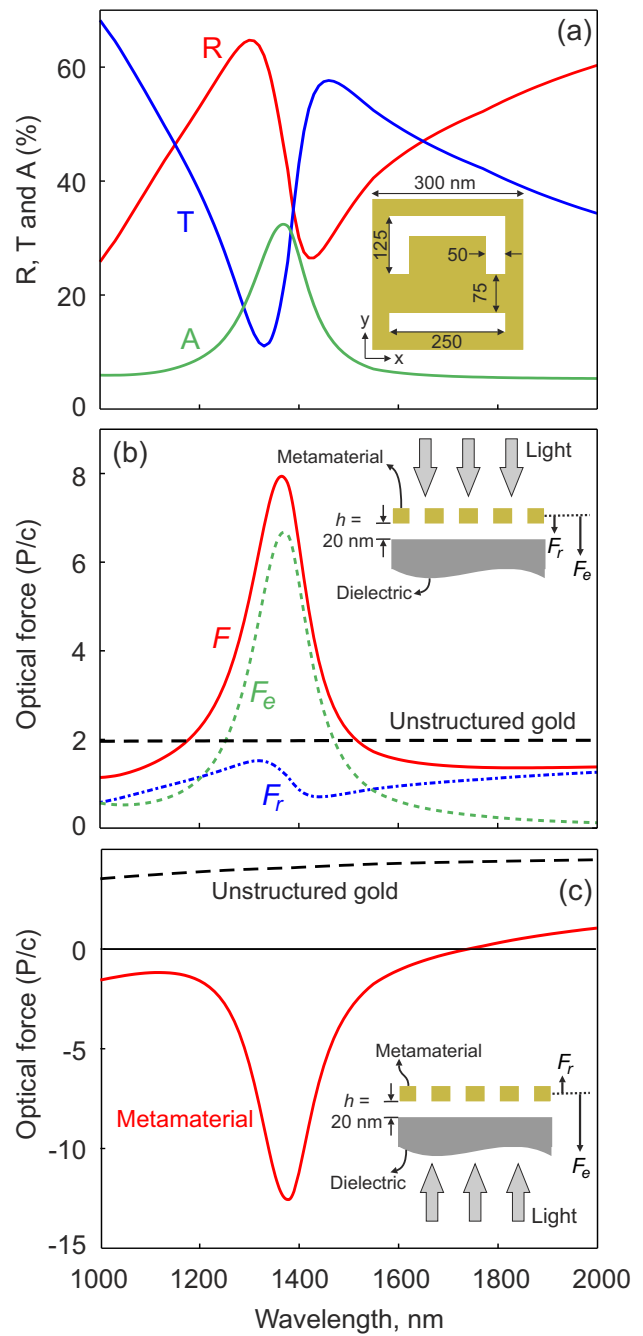}
\caption{\textbf{Optical forces between a plasmonic metamaterial and a dielectric surface. a,} Reflection $R$, transmission $T$ and absorption $A$ spectra of a 50~nm thick gold metamaterial located at a distance $h$~=~20~nm from the surface of a dielectric with refractive index $n$~=~2.5 for y-polarized light normally incident from free space. The inset shows the metamaterial unit cell geometry. \textbf{b,} Total optical force $F$ acting on a metamaterial illuminated from free-space as illustrated inset. Dashed lines show the evanescent $F_e$ and light pressure $F_r$ components of $F$. \textbf{c,} Total optical force acting on a metamaterial illuminated through the dielectric. For comparison, the total force acting on an \emph{unstructured} gold film in place of the metamaterial is shown in both \textbf{b} and \textbf{c}. In all cases, positive values denote forces acting in the direction of incident light propagation.}
\vspace{-10pt}
\end{figure}

For a plasmonic metamaterial both the radiation pressure and near-field forces are resonant (Fig~2b). The dispersion of the radiation pressure force $F_r$ is linked to variations in the metamaterial's absorbtion and reflection coefficients (Fig.~2a) and has a local maximum at a wavelength of 1320~nm corresponding to the reflectivity peak. When the structure is illuminated from free space the radiation pressure $F_r$ and near-field evanescent $F_e$ forces act in the same direction to reduce the gap $h$ between the metamaterial and the dielectric surface. The evanescent force is resonant at 1370~nm, the wavelength of the absorption peak, and exceeds the radiation force across the  entire spectral range from 1210 to 1550~nm, reaching a peak magnitude of approximately 6.7$P/c$ as compared to only 1.2$P/c$ for the radiation force. When light impinges on the sample through the transparent dielectric (Fig.~2c) the evanescent force acts \emph{in opposition} to the radiation pressure and again acts to reduce $h$. In this configuration, if the gold film were unstructured the total light force would push it away from the dielectric surface, but for a metamaterial the attractive near-field force is dominant and pulls the metamaterial film towards the dielectric.

The near-field force is related to the non-propagating evanescent field of localized plasmons in the nanostructure, which possesses no momentum. As such it does not contribute directly to the momentum balance with incident light; it simply adds to or subtracts from the radiation pressure force depending on the direction of the later. The evanescent force works to change the energy trapped in the nanostructure when the metamaterial is brought into close proximity with a surface. As the eigenfrequencies of a metamaterial are always red-shifted relative to their free-space values when the structure is close to or embedded in another medium with a refractive index exceeding unity, the closest possible presence of such a medium is energetically preferable. As a consequence, the resonant evanescent force between a metamaterial and nearby surface is always attractive.

Both the radiation pressure and evanescent forces are stronger when the metamaterial is illuminated through the dielectric than when light is incident on the structure from free space. In particular, higher values of $|F_e|$ (stronger evanescent fields between metamaterial and dielectric) are a consequence the metamaterial's higher absorption coefficient under illumination through the dielectric~\cite{Bliokh2008} (the increase in $|F_e|$ being proportional to that in $A$).

Figure~3 shows the dependence of total optical force on dielectric refractive index $n$ and gap size $h$ for free space illumination (as in Fig.~2b). As the refractive index of the dielectric increases from $n=1$ to 4, the magnitude of the optical force increases and the resonance red-shifts. A similar trend is seen with decreasing gap size: at $h$~=~5~nm the evanescent force is 33 times stronger than the radiation pressure. Both trends reflect the strong influence of near-field environment on the metamaterial resonances.

\begin{figure}
\includegraphics[width=82mm]{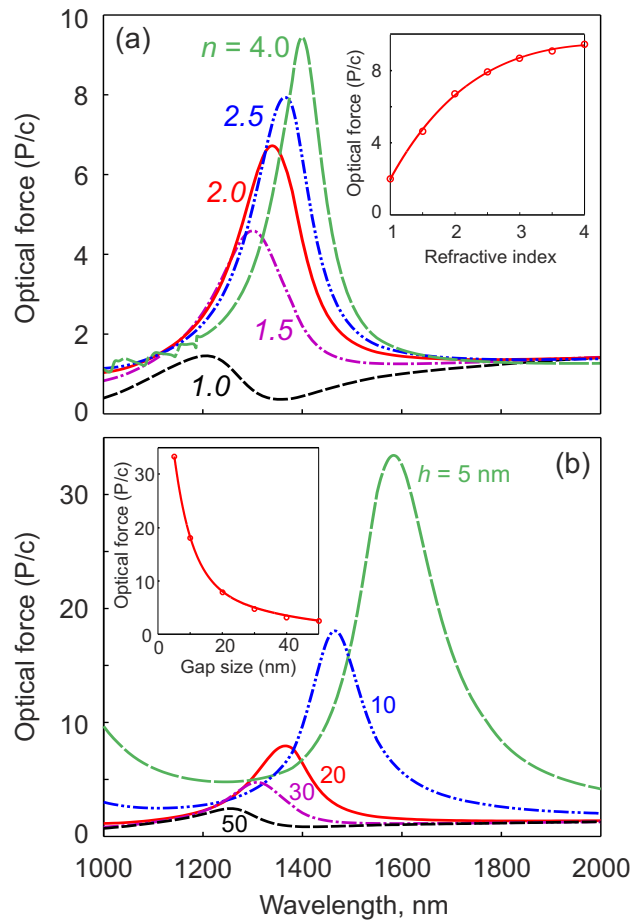}
\caption{\textbf{Optical forces between a plasmonic metamaterial and a dielectric surface.} Spectral dispersion of the total optical force (under illumination from free space as illustrated inset to Fig.~2b) for different values, as labeled, of: \textbf{a,} dielectric refractive index $n$ (at h~=~20~nm); \textbf{b,} gap size $h$ ($n$~=~2.5). Insets show peak optical force as a function of $n$ and $h$ respectively.}
\vspace{-10pt}
\end{figure}

Near-field optical forces are even greater at metallic surfaces. In this case we consider a metamaterial comprising a square array of rectangular slots (see inset to Fig.~4a). With a metallic backplane this structure supports a `magnetic resonance' (where anti-asymmetric currents are excited in the metamaterial and backplane) at which light is trapped (see inset to Fig.~4b) and strongly absorbed. Indeed, similar systems have previously been analyzed for `perfect' absorption applications\cite{Liu2010a,Liu2010,Hu2009}. Figure~4 shows the normal incidence reflection and absorption spectra of the structure and the dispersion of the optical forces acting on the metamaterial film. In this case the magnitude of the total optical force reaches $\sim$50$P/c$ at the absorption peak around 1095~nm.

\begin{figure}
\includegraphics[width=82mm]{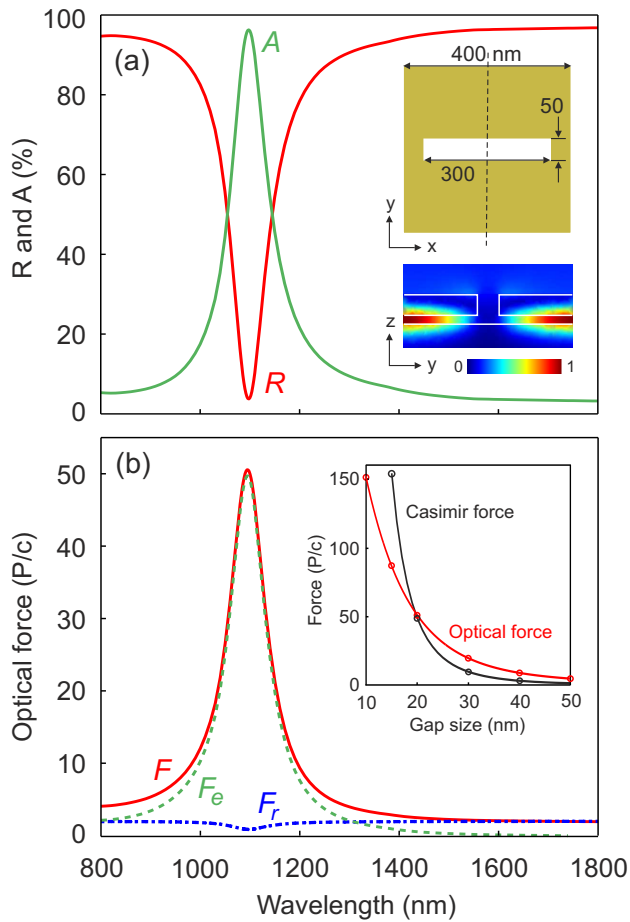}
\caption{\textbf{Optical forces between a plasmonic metamaterial and a metallic surface. a,} Normal incidence reflection $R$ and absorption $A$ spectra of a 50~nm thick gold metamaterial located at a distance $h$~=~20~nm from a gold surface (incident light propagation in the $-z$ direction as defined inset). The insets show the metamaterial unit cell geometry and a map of the normalized magnetic field intensity distribution at the 1095~nm resonance wavelength for a cross-section along the dashed line in the y-z plane. \textbf{b,} Evanescent $F_e$, radiation pressure $F_r$ and total optical force $F$ acting on the metamaterial. The inset shows peak optical force as a function of gap size $h$ and the corresponding dependence on $h$ of the Casimir force between two perfectly conducting plates (scaled assuming $I$ = 50~mW/$\mu$m$^2$).}
\vspace{-10pt}
\end{figure}

It is interesting to compare the enhanced optical forces between a metamaterial and metallic-surface with the force of gravity on the metamaterial film. The gravitational force on a 50~nm thick gold film is of order $1\times10^{-14}$~N/$\mu$m$^2$. An optical force of the same magnitude can be achieved between a metallic (gold) surface and a metamaterial 20~nm away when the metamaterial is illuminated at the 1095~nm resonant wavelength at an intensity of around $60$~nW/$\mu$m$^2$. Under such conditions the optical force will be sufficient to pull a metamaterial against gravity towards a surface. At shorter distances the pulling force becomes even stronger but other micro-/nanoscale forces also become important. The Casimir force $F_{c}=-(\pi^{2} \hbar c)/(240 d^{4})$ between two perfectly conducting plates separated by 20~nm is (at $\sim$8~nN/$\mu$m$^2$) equivalent to the optical force achieved at the same separation between a gold metamaterial and planar gold surface for an illumination intensity $I$ of approximately $50$~mW/$\mu$m$^2$ (see inset to Fig.~4b). However, for a perforated real metal thin film the Casimir force may easily be an order of magnitude smaller than the above estimate~\cite{Svetovoy2008,Rodriguez2011a,Lisanti2005}, in which case it would be surpassed by the near-field optical force at much lower intensities.

Indeed, as the Casimir and near-field forces depend differently on distance: $F_c \propto h^{-4}$ (or $h^{-3}$ for realistic metallic plates at values of $h$ smaller than the metal's plasma wavelength~\cite{Rodriguez2011a}) while $F_e \propto I~e^{-h/a}$ (where $a$ is the characteristic dimension of the nanostructural pattern), above a threshold value of intensity $I$ there will be a range of distances $h$ where the near-field force is dominant, and this range will broaden with increasing intensity. As a result, the near-field force is likely to prevail where conditions such as surface roughness limit the proximity of two objects.

With regard to the experimental observation of the evanescent force described here, photonic metamaterials can readily tolerate mW/$\mu$m$^2$ illumination intensities whereunder $F_e$ would exceed the gravitational force by several orders of magnitude. The great advantage of this near-field force for nanoscale manipulation is that it depends on both light intensity and wavelength, thereby offering dynamic controllability and spectral selectivity. Such forces could manifest themselves in a variety of resonant structural configurations including, for instance, single meta-molecules or plasmonic elements and may serve applications in optical trapping/tweezing and in the control of light with light via optically reconfigurable metamaterials. For example: an optical fibre scanning probe tip capped with a plasmonic metamaterial may be employed to pick up and re-position individual nano-objects; an array of optically switchable adhesion elements may dynamically control patterns of particles/flakes; optically tuning the position of a metamaterial elastically suspended near a metal surface (varying illumination intensity to adjust the balance between near-field and mechanical forces) will deliver broadband changes in the reflectivity of the structure.

In summary, we have demonstrated that the attractive near-field optical forces between a nanostructured plasmonic metamaterial film and a dielectric or metal surface can greatly exceed the associated radiation pressure and outstrip the Casimir forces. This force is sufficient to overcome gravity at illumination intensities of just a few tens of nW/$\mathrm{\mu}$m$^2$.\\

\begin{acknowledgments}
This work was supported by the Engineering and Physical Sciences Research Council [Project EP/G060363/1], The Royal Society (NIZ), and the China Scholarship Council (JZ).
\end{acknowledgments}


\begin{thebibliography}{36}
\expandafter\ifx\csname natexlab\endcsname\relax\def\natexlab#1{#1}\fi
\expandafter\ifx\csname bibnamefont\endcsname\relax
  \def\bibnamefont#1{#1}\fi
\expandafter\ifx\csname bibfnamefont\endcsname\relax
  \def\bibfnamefont#1{#1}\fi
\expandafter\ifx\csname citenamefont\endcsname\relax
  \def\citenamefont#1{#1}\fi
\expandafter\ifx\csname url\endcsname\relax
  \def\url#1{\texttt{#1}}\fi
\expandafter\ifx\csname urlprefix\endcsname\relax\def\urlprefix{URL }\fi
\providecommand{\bibinfo}[2]{#2}
\providecommand{\eprint}[2][]{\url{#2}}

\bibitem[{\citenamefont{Dholakia and \v{C}i\v{z}m\'ar}(2011)}]{Dholakia2011}
\bibinfo{author}{\bibfnamefont{K.}~\bibnamefont{Dholakia}} \bibnamefont{and}
  \bibinfo{author}{\bibfnamefont{T.}~\bibnamefont{\v{C}i\v{z}m\'ar}},
  \bibinfo{journal}{Nat. Photon.} \textbf{\bibinfo{volume}{5}},
  \bibinfo{pages}{335} (\bibinfo{year}{2011}).

\bibitem[{\citenamefont{Padgett and Bowman}(2011)}]{Padgett2011}
\bibinfo{author}{\bibfnamefont{M.}~\bibnamefont{Padgett}} \bibnamefont{and}
  \bibinfo{author}{\bibfnamefont{R.}~\bibnamefont{Bowman}},
  \bibinfo{journal}{Nat. Photon.} \textbf{\bibinfo{volume}{5}},
  \bibinfo{pages}{343} (\bibinfo{year}{2011}).

\bibitem[{\citenamefont{Dholakia and Zem\'{a}nek}(2010)}]{Dholakia2010}
\bibinfo{author}{\bibfnamefont{K.}~\bibnamefont{Dholakia}} \bibnamefont{and}
  \bibinfo{author}{\bibfnamefont{P.}~\bibnamefont{Zem\'{a}nek}},
  \bibinfo{journal}{Rev. Mod. Phys.} \textbf{\bibinfo{volume}{82}},
  \bibinfo{pages}{1767} (\bibinfo{year}{2010}).

\bibitem[{\citenamefont{Chen et~al.}(2011)\citenamefont{Chen, Ng, Lin, and
  Chan}}]{Chen2011a}
\bibinfo{author}{\bibfnamefont{J.}~\bibnamefont{Chen}},
  \bibinfo{author}{\bibfnamefont{J.}~\bibnamefont{Ng}},
  \bibinfo{author}{\bibfnamefont{Z.}~\bibnamefont{Lin}}, \bibnamefont{and}
  \bibinfo{author}{\bibfnamefont{C.~T.} \bibnamefont{Chan}},
  \bibinfo{journal}{Nat. Photon.} \textbf{\bibinfo{volume}{5}},
  \bibinfo{pages}{531} (\bibinfo{year}{2011}).

\bibitem[{\citenamefont{Kippenberg and Vahala}(2008)}]{Kippenberg2008}
\bibinfo{author}{\bibfnamefont{T.~J.} \bibnamefont{Kippenberg}}
  \bibnamefont{and} \bibinfo{author}{\bibfnamefont{K.~J.}
  \bibnamefont{Vahala}}, \bibinfo{journal}{Science}
  \textbf{\bibinfo{volume}{321}}, \bibinfo{pages}{1172} (\bibinfo{year}{2008}).

\bibitem[{\citenamefont{Li et~al.}(2008)\citenamefont{Li, Pernice, Xiong,
  Baehr-Jones, Hochberg, and Tang}}]{Li2008a}
\bibinfo{author}{\bibfnamefont{M.}~\bibnamefont{Li}},
  \bibinfo{author}{\bibfnamefont{W.~H.~P.} \bibnamefont{Pernice}},
  \bibinfo{author}{\bibfnamefont{C.}~\bibnamefont{Xiong}},
  \bibinfo{author}{\bibfnamefont{T.}~\bibnamefont{Baehr-Jones}},
  \bibinfo{author}{\bibfnamefont{M.}~\bibnamefont{Hochberg}}, \bibnamefont{and}
  \bibinfo{author}{\bibfnamefont{H.~X.} \bibnamefont{Tang}},
  \bibinfo{journal}{Nature} \textbf{\bibinfo{volume}{456}},
  \bibinfo{pages}{480} (\bibinfo{year}{2008}).

\bibitem[{\citenamefont{Ma and Povinelli}(2011)}]{Ma2011}
\bibinfo{author}{\bibfnamefont{J.}~\bibnamefont{Ma}} \bibnamefont{and}
  \bibinfo{author}{\bibfnamefont{M.~L.} \bibnamefont{Povinelli}},
  \bibinfo{journal}{Opt. Exp.} \textbf{\bibinfo{volume}{19}},
  \bibinfo{pages}{10102} (\bibinfo{year}{2011}).

\bibitem[{\citenamefont{Roels et~al.}(2009)\citenamefont{Roels, De~Vlaminck,
  Lagae, Maes, Van~Thourhout, and Baets}}]{Roels2009}
\bibinfo{author}{\bibfnamefont{J.}~\bibnamefont{Roels}},
  \bibinfo{author}{\bibfnamefont{I.}~\bibnamefont{De~Vlaminck}},
  \bibinfo{author}{\bibfnamefont{L.}~\bibnamefont{Lagae}},
  \bibinfo{author}{\bibfnamefont{B.}~\bibnamefont{Maes}},
  \bibinfo{author}{\bibfnamefont{D.}~\bibnamefont{Van~Thourhout}},
  \bibnamefont{and} \bibinfo{author}{\bibfnamefont{R.}~\bibnamefont{Baets}},
  \bibinfo{journal}{Nat. Nanotech.} \textbf{\bibinfo{volume}{4}},
  \bibinfo{pages}{510} (\bibinfo{year}{2009}).

\bibitem[{\citenamefont{Wiederhecker et~al.}(2009)\citenamefont{Wiederhecker,
  Chen, Gondarenko, and Lipson}}]{Wiederhecker2009}
\bibinfo{author}{\bibfnamefont{G.~S.} \bibnamefont{Wiederhecker}},
  \bibinfo{author}{\bibfnamefont{L.}~\bibnamefont{Chen}},
  \bibinfo{author}{\bibfnamefont{A.}~\bibnamefont{Gondarenko}},
  \bibnamefont{and} \bibinfo{author}{\bibfnamefont{M.}~\bibnamefont{Lipson}},
  \bibinfo{journal}{Nature} \textbf{\bibinfo{volume}{462}},
  \bibinfo{pages}{633} (\bibinfo{year}{2009}).

\bibitem[{\citenamefont{Taniyama et~al.}(2008)\citenamefont{Taniyama, Notomi,
  Kuramochi, Yamamoto, Yoshikawa, Torii, and Kuga}}]{Taniyama2008}
\bibinfo{author}{\bibfnamefont{H.}~\bibnamefont{Taniyama}},
  \bibinfo{author}{\bibfnamefont{M.}~\bibnamefont{Notomi}},
  \bibinfo{author}{\bibfnamefont{E.}~\bibnamefont{Kuramochi}},
  \bibinfo{author}{\bibfnamefont{T.}~\bibnamefont{Yamamoto}},
  \bibinfo{author}{\bibfnamefont{Y.}~\bibnamefont{Yoshikawa}},
  \bibinfo{author}{\bibfnamefont{Y.}~\bibnamefont{Torii}}, \bibnamefont{and}
  \bibinfo{author}{\bibfnamefont{T.}~\bibnamefont{Kuga}},
  \bibinfo{journal}{Phys. Rev. B} \textbf{\bibinfo{volume}{78}},
  \bibinfo{pages}{165129} (\bibinfo{year}{2008}).

\bibitem[{\citenamefont{Liu et~al.}(2009)\citenamefont{Liu, Povinelli, and
  Fan}}]{Liu2009}
\bibinfo{author}{\bibfnamefont{V.}~\bibnamefont{Liu}},
  \bibinfo{author}{\bibfnamefont{M.}~\bibnamefont{Povinelli}},
  \bibnamefont{and} \bibinfo{author}{\bibfnamefont{S.}~\bibnamefont{Fan}},
  \bibinfo{journal}{Opt. Exp.} \textbf{\bibinfo{volume}{17}},
  \bibinfo{pages}{21897} (\bibinfo{year}{2009}).

\bibitem[{\citenamefont{Munday et~al.}(2009)\citenamefont{Munday, Capasso, and
  Parsegian}}]{Munday2009}
\bibinfo{author}{\bibfnamefont{J.~N.} \bibnamefont{Munday}},
  \bibinfo{author}{\bibfnamefont{F.}~\bibnamefont{Capasso}}, \bibnamefont{and}
  \bibinfo{author}{\bibfnamefont{V.~A.} \bibnamefont{Parsegian}},
  \bibinfo{journal}{Nature} \textbf{\bibinfo{volume}{457}},
  \bibinfo{pages}{170} (\bibinfo{year}{2009}).

\bibitem[{\citenamefont{Lebedev}(1901)}]{Lebedev1901}
\bibinfo{author}{\bibfnamefont{P.}~\bibnamefont{Lebedev}},
  \bibinfo{journal}{Ann. Phys.} \textbf{\bibinfo{volume}{311}},
  \bibinfo{pages}{433} (\bibinfo{year}{1901}).

\bibitem[{\citenamefont{Nichols and Hull}(1903)}]{Nichols1903}
\bibinfo{author}{\bibfnamefont{E.~F.} \bibnamefont{Nichols}} \bibnamefont{and}
  \bibinfo{author}{\bibfnamefont{G.~F.} \bibnamefont{Hull}},
  \bibinfo{journal}{Phys. Rev.} \textbf{\bibinfo{volume}{17}},
  \bibinfo{pages}{26} (\bibinfo{year}{1903}).

\bibitem[{\citenamefont{Swartzlander~Jr.
  et~al.}(2011)\citenamefont{Swartzlander~Jr., Peterson, Artusio-Glimpse, and
  Raisanen}}]{SwartzlanderJr.2011}
\bibinfo{author}{\bibfnamefont{G.~A.} \bibnamefont{Swartzlander~Jr.}},
  \bibinfo{author}{\bibfnamefont{T.~J.} \bibnamefont{Peterson}},
  \bibinfo{author}{\bibfnamefont{A.~B.} \bibnamefont{Artusio-Glimpse}},
  \bibnamefont{and} \bibinfo{author}{\bibfnamefont{A.~D.}
  \bibnamefont{Raisanen}}, \bibinfo{journal}{Nat. Photon.}
  \textbf{\bibinfo{volume}{5}}, \bibinfo{pages}{48} (\bibinfo{year}{2011}).

\bibitem[{\citenamefont{Autumn et~al.}(2002)\citenamefont{Autumn, Sitti, Liang,
  Peattie, Hansen, Sponberg, Kenny, Fearing, Israelachvili, and
  Full}}]{Autumn2002}
\bibinfo{author}{\bibfnamefont{K.}~\bibnamefont{Autumn}},
  \bibinfo{author}{\bibfnamefont{M.}~\bibnamefont{Sitti}},
  \bibinfo{author}{\bibfnamefont{Y.~A.} \bibnamefont{Liang}},
  \bibinfo{author}{\bibfnamefont{A.~M.} \bibnamefont{Peattie}},
  \bibinfo{author}{\bibfnamefont{W.~R.} \bibnamefont{Hansen}},
  \bibinfo{author}{\bibfnamefont{S.}~\bibnamefont{Sponberg}},
  \bibinfo{author}{\bibfnamefont{T.~W.} \bibnamefont{Kenny}},
  \bibinfo{author}{\bibfnamefont{R.}~\bibnamefont{Fearing}},
  \bibinfo{author}{\bibfnamefont{J.~N.} \bibnamefont{Israelachvili}},
  \bibnamefont{and} \bibinfo{author}{\bibfnamefont{R.~J.} \bibnamefont{Full}},
  \bibinfo{journal}{P. Natl. Acad. Sci. USA} \textbf{\bibinfo{volume}{99}},
  \bibinfo{pages}{12252} (\bibinfo{year}{2002}).

\bibitem[{\citenamefont{Zhao et~al.}(2010)\citenamefont{Zhao, Tassin, Koschny,
  and Soukoulis}}]{Zhao2010}
\bibinfo{author}{\bibfnamefont{R.}~\bibnamefont{Zhao}},
  \bibinfo{author}{\bibfnamefont{P.}~\bibnamefont{Tassin}},
  \bibinfo{author}{\bibfnamefont{T.}~\bibnamefont{Koschny}}, \bibnamefont{and}
  \bibinfo{author}{\bibfnamefont{C.~M.} \bibnamefont{Soukoulis}},
  \bibinfo{journal}{Opt. Exp.} \textbf{\bibinfo{volume}{18}},
  \bibinfo{pages}{25665} (\bibinfo{year}{2010}).

\bibitem[{\citenamefont{Liu et~al.}(2010{\natexlab{a}})\citenamefont{Liu,
  Zentgraf, Liu, Bartal, and Zhang}}]{Liu2010b}
\bibinfo{author}{\bibfnamefont{M.}~\bibnamefont{Liu}},
  \bibinfo{author}{\bibfnamefont{T.}~\bibnamefont{Zentgraf}},
  \bibinfo{author}{\bibfnamefont{Y.}~\bibnamefont{Liu}},
  \bibinfo{author}{\bibfnamefont{G.}~\bibnamefont{Bartal}}, \bibnamefont{and}
  \bibinfo{author}{\bibfnamefont{X.}~\bibnamefont{Zhang}},
  \bibinfo{journal}{Nat. Nanotech.} \textbf{\bibinfo{volume}{5}},
  \bibinfo{pages}{570} (\bibinfo{year}{2010}{\natexlab{a}}).

\bibitem[{\citenamefont{Wang et~al.}(2011)\citenamefont{Wang, Ng, Liu, Zheng,
  Hang, and Chan}}]{Wang2011}
\bibinfo{author}{\bibfnamefont{S.~B.} \bibnamefont{Wang}},
  \bibinfo{author}{\bibfnamefont{J.}~\bibnamefont{Ng}},
  \bibinfo{author}{\bibfnamefont{H.}~\bibnamefont{Liu}},
  \bibinfo{author}{\bibfnamefont{H.~H.} \bibnamefont{Zheng}},
  \bibinfo{author}{\bibfnamefont{Z.~H.} \bibnamefont{Hang}}, \bibnamefont{and}
  \bibinfo{author}{\bibfnamefont{C.~T.} \bibnamefont{Chan}},
  \bibinfo{journal}{Phys. Rev. B} \textbf{\bibinfo{volume}{84}},
  \bibinfo{pages}{075114} (\bibinfo{year}{2011}).

\bibitem[{\citenamefont{Liu et~al.}(2011)\citenamefont{Liu, Ng, Wang, Lin,
  Hang, Chan, and Zhu}}]{Liu2011}
\bibinfo{author}{\bibfnamefont{H.}~\bibnamefont{Liu}},
  \bibinfo{author}{\bibfnamefont{J.}~\bibnamefont{Ng}},
  \bibinfo{author}{\bibfnamefont{S.~B.} \bibnamefont{Wang}},
  \bibinfo{author}{\bibfnamefont{Z.~F.} \bibnamefont{Lin}},
  \bibinfo{author}{\bibfnamefont{Z.~H.} \bibnamefont{Hang}},
  \bibinfo{author}{\bibfnamefont{C.~T.} \bibnamefont{Chan}}, \bibnamefont{and}
  \bibinfo{author}{\bibfnamefont{S.~N.} \bibnamefont{Zhu}},
  \bibinfo{journal}{Phys. Rev. Lett.} \textbf{\bibinfo{volume}{106}},
  \bibinfo{pages}{087401} (\bibinfo{year}{2011}).

\bibitem[{\citenamefont{Juan et~al.}(2011)\citenamefont{Juan, Righini, and
  Quidant}}]{Juan2011}
\bibinfo{author}{\bibfnamefont{M.~L.} \bibnamefont{Juan}},
  \bibinfo{author}{\bibfnamefont{M.}~\bibnamefont{Righini}}, \bibnamefont{and}
  \bibinfo{author}{\bibfnamefont{R.}~\bibnamefont{Quidant}},
  \bibinfo{journal}{Nat. Photon.} \textbf{\bibinfo{volume}{5}},
  \bibinfo{pages}{349} (\bibinfo{year}{2011}).

\bibitem[{\citenamefont{Veselago}(1968)}]{Veselago1968}
\bibinfo{author}{\bibfnamefont{V.~G.} \bibnamefont{Veselago}},
  \bibinfo{journal}{Sov. Phys. Usp.} \textbf{\bibinfo{volume}{10}},
  \bibinfo{pages}{509} (\bibinfo{year}{1968}).

\bibitem[{\citenamefont{Lezec et~al.}(2011)\citenamefont{Lezec, Agarwal, and
  Chau}}]{Lezec2011}
\bibinfo{author}{\bibfnamefont{H.~J.} \bibnamefont{Lezec}},
  \bibinfo{author}{\bibfnamefont{A.}~\bibnamefont{Agarwal}}, \bibnamefont{and}
  \bibinfo{author}{\bibfnamefont{K.}~\bibnamefont{Chau}}, in
  \emph{\bibinfo{booktitle}{NanoMeta 2011}}
  (\bibinfo{address}{Seefeld-in-Tirol, Austria}, \bibinfo{year}{2011}).

\bibitem[{\citenamefont{T{\o}rrissen}()}]{Torrissen}
\bibinfo{author}{\bibfnamefont{B.~C.} \bibnamefont{T{\o}rrissen}}
  [\bibinfo{url}{http:$//$en.wikipedia.org$/$wiki$/$File:Gecko $\textunderscore$foot$\textunderscore$on$\textunderscore$glass.JPG}]
  (\bibinfo{publisher}{CC BY-SA 3.0}).

\bibitem[{\citenamefont{Rodriguez
  et~al.}(2011{\natexlab{a}})\citenamefont{Rodriguez, McCauley, Hui, Woolf,
  Iwase, Capasso, Lon\v{c}ar, and Johnson}}]{Rodriguez2011}
\bibinfo{author}{\bibfnamefont{A.~W.} \bibnamefont{Rodriguez}},
  \bibinfo{author}{\bibfnamefont{A.~P.} \bibnamefont{McCauley}},
  \bibinfo{author}{\bibfnamefont{P.~C.} \bibnamefont{Hui}},
  \bibinfo{author}{\bibfnamefont{D.}~\bibnamefont{Woolf}},
  \bibinfo{author}{\bibfnamefont{E.}~\bibnamefont{Iwase}},
  \bibinfo{author}{\bibfnamefont{F.}~\bibnamefont{Capasso}},
  \bibinfo{author}{\bibfnamefont{M.}~\bibnamefont{Lon\v{c}ar}},
  \bibnamefont{and} \bibinfo{author}{\bibfnamefont{S.~G.}
  \bibnamefont{Johnson}}, \bibinfo{journal}{Opt. Exp.}
  \textbf{\bibinfo{volume}{19}}, \bibinfo{pages}{2225}
  (\bibinfo{year}{2011}{\natexlab{a}}).

\bibitem[{\citenamefont{Palik}(1984)}]{Palik1984}
\bibinfo{editor}{\bibfnamefont{E.~D.} \bibnamefont{Palik}}, ed.,
  \emph{\bibinfo{title}{Handbook of Optical Constants of Solids}}
  (\bibinfo{publisher}{Academic Press}, \bibinfo{address}{Orlando},
  \bibinfo{year}{1984}).

\bibitem[{\citenamefont{Povinelli et~al.}(2005)\citenamefont{Povinelli,
  Johnson, Lon\v{c}ar, Ibanescu, Smythe, Capasso, and
  Joannopoulos}}]{Povinelli2005}
\bibinfo{author}{\bibfnamefont{M.~L.} \bibnamefont{Povinelli}},
  \bibinfo{author}{\bibfnamefont{S.~G.} \bibnamefont{Johnson}},
  \bibinfo{author}{\bibfnamefont{M.}~\bibnamefont{Lon\v{c}ar}},
  \bibinfo{author}{\bibfnamefont{M.}~\bibnamefont{Ibanescu}},
  \bibinfo{author}{\bibfnamefont{E.~J.} \bibnamefont{Smythe}},
  \bibinfo{author}{\bibfnamefont{F.}~\bibnamefont{Capasso}}, \bibnamefont{and}
  \bibinfo{author}{\bibfnamefont{J.~D.} \bibnamefont{Joannopoulos}},
  \bibinfo{journal}{Opt. Exp.} \textbf{\bibinfo{volume}{13}},
  \bibinfo{pages}{8286} (\bibinfo{year}{2005}).

\bibitem[{\citenamefont{Tanaka et~al.}(2010)\citenamefont{Tanaka, Plum, Ou,
  Uchino, and Zheludev}}]{Tanaka2010}
\bibinfo{author}{\bibfnamefont{K.}~\bibnamefont{Tanaka}},
  \bibinfo{author}{\bibfnamefont{E.}~\bibnamefont{Plum}},
  \bibinfo{author}{\bibfnamefont{J.~Y.} \bibnamefont{Ou}},
  \bibinfo{author}{\bibfnamefont{T.}~\bibnamefont{Uchino}}, \bibnamefont{and}
  \bibinfo{author}{\bibfnamefont{N.~I.} \bibnamefont{Zheludev}},
  \bibinfo{journal}{Phys. Rev. Lett.} \textbf{\bibinfo{volume}{105}},
  \bibinfo{pages}{227403} (\bibinfo{year}{2010}).

\bibitem[{\citenamefont{Nikolaenko et~al.}(2010)\citenamefont{Nikolaenko,
  De~Angelis, Boden, Papasimakis, Ashburn, Di~Fabrizio, and
  Zheludev}}]{Nikolaenko2010}
\bibinfo{author}{\bibfnamefont{A.}~\bibnamefont{Nikolaenko}},
  \bibinfo{author}{\bibfnamefont{F.}~\bibnamefont{De~Angelis}},
  \bibinfo{author}{\bibfnamefont{S.~A.} \bibnamefont{Boden}},
  \bibinfo{author}{\bibfnamefont{N.}~\bibnamefont{Papasimakis}},
  \bibinfo{author}{\bibfnamefont{P.}~\bibnamefont{Ashburn}},
  \bibinfo{author}{\bibfnamefont{E.}~\bibnamefont{Di~Fabrizio}},
  \bibnamefont{and} \bibinfo{author}{\bibfnamefont{N.~I.}
  \bibnamefont{Zheludev}}, \bibinfo{journal}{Phys. Rev. Lett.}
  \textbf{\bibinfo{volume}{104}}, \bibinfo{pages}{153902}
  (\bibinfo{year}{2010}).

\bibitem[{\citenamefont{Bliokh et~al.}(2008)\citenamefont{Bliokh, Bliokh,
  Freilikher, Savel'ev, and Nori}}]{Bliokh2008}
\bibinfo{author}{\bibfnamefont{K.~Y.} \bibnamefont{Bliokh}},
  \bibinfo{author}{\bibfnamefont{Y.~P.} \bibnamefont{Bliokh}},
  \bibinfo{author}{\bibfnamefont{V.}~\bibnamefont{Freilikher}},
  \bibinfo{author}{\bibfnamefont{S.}~\bibnamefont{Savel'ev}}, \bibnamefont{and}
  \bibinfo{author}{\bibfnamefont{F.}~\bibnamefont{Nori}},
  \bibinfo{journal}{Rev. Mod. Phys.} \textbf{\bibinfo{volume}{80}},
  \bibinfo{pages}{1201} (\bibinfo{year}{2008}).

\bibitem[{\citenamefont{Liu et~al.}(2010{\natexlab{b}})\citenamefont{Liu,
  Mensch, Weiss, Hentschel, and Giessen}}]{Liu2010a}
\bibinfo{author}{\bibfnamefont{N.}~\bibnamefont{Liu}},
  \bibinfo{author}{\bibfnamefont{M.}~\bibnamefont{Mensch}},
  \bibinfo{author}{\bibfnamefont{T.}~\bibnamefont{Weiss}},
  \bibinfo{author}{\bibfnamefont{M.}~\bibnamefont{Hentschel}},
  \bibnamefont{and} \bibinfo{author}{\bibfnamefont{H.}~\bibnamefont{Giessen}},
  \bibinfo{journal}{Nano Lett.} \textbf{\bibinfo{volume}{10}},
  \bibinfo{pages}{2342–2348} (\bibinfo{year}{2010}{\natexlab{b}}).

\bibitem[{\citenamefont{Liu et~al.}(2010{\natexlab{c}})\citenamefont{Liu,
  Starr, Starr, and Padilla}}]{Liu2010}
\bibinfo{author}{\bibfnamefont{X.}~\bibnamefont{Liu}},
  \bibinfo{author}{\bibfnamefont{T.}~\bibnamefont{Starr}},
  \bibinfo{author}{\bibfnamefont{A.~F.} \bibnamefont{Starr}}, \bibnamefont{and}
  \bibinfo{author}{\bibfnamefont{W.~J.} \bibnamefont{Padilla}},
  \bibinfo{journal}{Phys. Rev. Lett.} \textbf{\bibinfo{volume}{104}},
  \bibinfo{pages}{207403} (\bibinfo{year}{2010}{\natexlab{c}}).

\bibitem[{\citenamefont{Hu et~al.}(2009)\citenamefont{Hu, Zhao, Chen, and
  Luo}}]{Hu2009}
\bibinfo{author}{\bibfnamefont{C.}~\bibnamefont{Hu}},
  \bibinfo{author}{\bibfnamefont{Z.}~\bibnamefont{Zhao}},
  \bibinfo{author}{\bibfnamefont{X.}~\bibnamefont{Chen}}, \bibnamefont{and}
  \bibinfo{author}{\bibfnamefont{X.}~\bibnamefont{Luo}}, \bibinfo{journal}{Opt.
  Exp.} \textbf{\bibinfo{volume}{17}}, \bibinfo{pages}{11039}
  (\bibinfo{year}{2009}).

\bibitem[{\citenamefont{Svetovoy et~al.}(2008)\citenamefont{Svetovoy, van Zwol,
  Palasantzas, and De~Hosson}}]{Svetovoy2008}
\bibinfo{author}{\bibfnamefont{V.~B.} \bibnamefont{Svetovoy}},
  \bibinfo{author}{\bibfnamefont{P.~J.} \bibnamefont{van Zwol}},
  \bibinfo{author}{\bibfnamefont{G.}~\bibnamefont{Palasantzas}},
  \bibnamefont{and} \bibinfo{author}{\bibfnamefont{J.~T.~M.}
  \bibnamefont{De~Hosson}}, \bibinfo{journal}{Phys. Rev. B}
  \textbf{\bibinfo{volume}{77}}, \bibinfo{pages}{035439}
  (\bibinfo{year}{2008}).

\bibitem[{\citenamefont{Rodriguez
  et~al.}(2011{\natexlab{b}})\citenamefont{Rodriguez, Capasso, and
  Johnson}}]{Rodriguez2011a}
\bibinfo{author}{\bibfnamefont{A.~W.} \bibnamefont{Rodriguez}},
  \bibinfo{author}{\bibfnamefont{F.}~\bibnamefont{Capasso}}, \bibnamefont{and}
  \bibinfo{author}{\bibfnamefont{S.~G.} \bibnamefont{Johnson}},
  \bibinfo{journal}{Nat. Photon.} \textbf{\bibinfo{volume}{5}},
  \bibinfo{pages}{211} (\bibinfo{year}{2011}{\natexlab{b}}).

\bibitem[{\citenamefont{Lisanti et~al.}(2005)\citenamefont{Lisanti, Iannuzzi,
  and Capasso}}]{Lisanti2005}
\bibinfo{author}{\bibfnamefont{M.}~\bibnamefont{Lisanti}},
  \bibinfo{author}{\bibfnamefont{D.}~\bibnamefont{Iannuzzi}}, \bibnamefont{and}
  \bibinfo{author}{\bibfnamefont{F.}~\bibnamefont{Capasso}},
  \bibinfo{journal}{P. Natl. Acad. Sci. USA} \textbf{\bibinfo{volume}{102}},
  \bibinfo{pages}{11989–11992} (\bibinfo{year}{2005}).

\end{thebibliography}

\end{document}